\documentclass[11pt]{article}
\newcommand{\be}{\begin{equation}}
\newcommand{\ee}{\end{equation}}
\newcommand{\ba}{\begin{array}}
\newcommand{\ea}{\end{array}}

\newcommand{\dys}{\displaystyle}
\newcommand{\qquer}{\overline{q}}
\newcommand{\Fquer}{\overline{F}}
\newcommand{\rom}{\rule{0mm}{0.5mm}}
\newcommand{\gl}{\! =\! }
\renewcommand{\theequation}{\arabic{equation}}

\begin{document}
\bibliographystyle{unsrt}

\newpage
\setcounter{page}{0}
\begin{titlepage}
\thispagestyle{empty}
  \begin{center}
    \vspace*{1cm}
{\Large \bf The modified XXZ Heisenberg chain, conformal invariance, \\[3mm]
surface exponents of \mbox{\boldmath $c < 1$} systems, and hidden \\[5mm]
symmetries of the finite chains \footnote{Lecture given by V. Rittenberg 
at the Conference on Yang-Baxter equations, conformal invariance and 
integrability, Canberra 1989} }\\[25mm] 
   
{\Large U. Grimm and V. Rittenberg }\\[10mm]

Physikalisches Institut der Universit\"at Bonn \\
Nu{\ss}allee 12, 5300 Bonn 1, West-Germany

\end{center}

\vspace{35mm}

\hspace{-\parindent}{\bf Abstract:}
The spin-$1/2$ XXZ Heisenberg chain with two types
of boundary terms is considered. For the first type,
the Hamiltonian is hermitian but not for the second type
which includes the $U_{q}[SU(2)]$ symmetric case.
It is shown that for a certain ``tuning'' between the
anisotropy angle and the boundary terms the spectra present
unexpected degeneracies. These degeneracies are related
to the structure of the irreducible representations of
the Virasoro algebras for $c<1$. 

\vfill

\centerline{\fbox{{\bf
\,Preprint version BONN-HE-90-01 of
{\em Int.\ J.\ Mod.\ Phys}.\/ B4 (1990) 969--978\,}}}
\vspace*{10mm}
\end{titlepage}

\newpage
\setcounter{equation}{0}
\pagestyle{plain}
 
Some time ago \cite{1} we have addressed ourselves the problem of universality
classes in the case of two-dimensional systems at a second-order phase
transition. Assuming that conformal invariance can be used (which is not 
always the case), one is tempted to define the universality class through
a certain modular invariant partition function corresponding to a given
central charge of the Virasoro algebra. In the case of the minimal series
for example, the modular invariant partition functions were given by
Cappelli et al. \cite{2} . A modular invariant partition function does not
however specify ``who is who'' in the list of primary fields. A given primary
operator however can be associated with a symmetry breaking operator
(corresponding to an order parameter) or to a thermal operator (its anomalous
dimension determines the specific heat). Phrased in a different way, the 
question is if two ore more systems defined on a lattice and having different
symmetries cannot correspond to the same modular invariant partition function.
We think that the answer is ``yes'' and that the complete description of the
system needs the knowledge of all the partition functions corresponding
to the different boundary conditions (BC) compatible with translational
invariance (all {\em toroidal BC}). In order to get an insight into 
the problem we have shown that one can project out from the Coulomb gas
partition functions, the partition functions of several systems, all of them
having the same modular invariant partition function. In this way, 
for the minimal series we have shown that for each given modular invariant
partition function we have four different systems and we called the 
corresponding series the $p\,$-state Potts, tricritical $p\,$-state Potts,
the $O(p)$, and the low temperature $O(p)$ models. A similar picture
was proposed earlier on quite different grounds by Nienhuis et al. \cite{3} .
Behind each of the series stay different short-distance algebras.
This work was extended later \cite{4} to the case of the $N\gl 1$ 
superconformal series where for each for each of the modular invariant
partition functions of Cappelli \cite{5} we have found two different models.
One remarkable fact is that when applied to the XXZ spin-$1/2$ Heisenberg
chain, for one class of models, the projection mechanism also works for finite
chains. A similar phenomenon occurs if the projection mechanism is applied
to the Fateev-Zamolodchikov spin-$1$ quantum chain \cite{4}.
 
In the present lecture we will describe the partition functions of the minimal
models described above in the case of {\em free BC}. As a by-product
we will show some ``miraculous'' properties of the XXZ chain with modified
BC. Some of the results presented here were already published \cite{6},
some others are new.
 
The modified XXZ model with $N$ sites is defined by the Hamiltonian
\be
H = - \frac{\gamma}{2\pi\sin \gamma} 
\left\{ \sum_{j=1}^{N-1} \left[ \sigma^{x}_{j} \sigma^{x}_{j+1} +
\sigma^{y}_{j} \sigma^{y}_{j+1} 
- \cos \gamma \sigma^{z}_{j}\sigma^{z}_{j+1} \right]
+ p \sigma^{z}_{1} + p^{\prime} \sigma^{z}_{N} \right\} \; \; ,
\label{e1} \ee
where $\sigma^{x}$, $\sigma^{y}$, and $\sigma^{z}$ are Pauli matrices
and $p$, $p^{\prime}$ describe the coupling to external fields.
The special case \mbox{$p\gl p^{\prime} \! = \! 0$} corresponds to 
free BC.
 
Instead of describing the anisotropy of the Hamiltonian through
the parameter $\gamma$ it is useful to define
\be
h = \frac{1}{4} \left( 1 - \frac{\gamma}{\pi} \right) ^{-1}
\vspace*{1cm} (h \geq \frac{1}{4} ) \; \; . 
\label{e2} \ee
As discussed in Ref.~\cite{7}, $h$ is related to the compactification
radius $R$ of the bosonic string \mbox{($h\gl R^{2}/2$)}. It also
is convenient to denote
\be \ba{cccc}
p = i \sin \gamma \coth \alpha & , & 
p^{\prime} = i \sin \gamma \coth \alpha^{\prime} & ,\ea 
\label{e3} \ee
where $\alpha$ and $\alpha^{\prime}$ are complex parameters. It turns
out that two parametrisations of $\alpha$ and $\alpha^{\prime}$ 
leave the spectrum real:
\renewcommand{\theequation}{4.\alph{equation}}
\setcounter{equation}{0}
\be \ba{cccc}
\alpha = -\frac{i \pi}{2} \psi + \frac{i \pi}{2} a & , &
\alpha^{\prime} = -\frac{i \pi}{2} \psi -\frac{i \pi}{2} a \ea
\label{e4a} \ee
where $\psi$ and $a$ are real (in this case the Hamiltonian is
hermitian), and
\be \ba{ccc}
\alpha = -\frac{i \pi}{2} \psi + \frac{\pi}{2} b & , &
\alpha^{\prime} = -\frac{i \pi}{2} \psi -\frac{\pi}{2} b \ea
\label{e4b} \ee
\renewcommand{\theequation}{\arabic{equation}}
\setcounter{equation}{4}
where $\psi$ and $b$ are real (in this case the Hamiltonian is 
{\em not} hermitian).
Note that for \mbox{$a\! = \! b \! = \! 0$}, \mbox{$\psi\gl 1$}
one recovers the case of free BC and that the two BC coincide
for \mbox{$a\! = \! b \! = \! 0$}. The case \mbox{$b \rightarrow \infty$}
for the BC (\ref{e4b}) is very special since in this case the Hamiltonian
has the quantum algebra \mbox{$U_{q}[SU(2)]$} with 
\mbox{$q\! = \! e^{i \gamma}$} as symmetry.
 
We notice that the charge operator
\be
\hat{Q} = \frac{1}{2} \sum_{j=1}^{N} \sigma^{z}_{j}
\label{e5} \ee
commutes with $H$ and that its eigenvalues $Q$ are integer (half-integer)
when $N$ is even (odd). Let $E_{Q;i}(N)$ be the energy levels,
\mbox{$i\! = \! 0,1,\ldots \;$}, in the charge sector $Q$ of the Hamiltonian
with $N$ sites and $E_{0}^{F}(N)$ the ground-state energy of the Hamiltonian
with free BC.
We consider the following quantities:
\be \ba{lcl}
\overline{E}_{Q;i} & = & \dys{\frac{N}{\pi} 
\left( E_{Q;i}(N) - E_{0}^{F}(N) \right)} \vspace{4mm} \\
{\cal E}_{Q}(N,z) & = & \dys{\sum_{i} z^{\overline{E}_{Q;i}(N)}} \vspace{4mm} \\
{\cal E}_{Q}(z) & = & \dys{\lim_{z \rightarrow \infty} {\cal E}_{Q}(N,z)}
\; \; . \ea
\label{e6} \ee
Using numerical estimates from chains up to $18$ sites as well as analytical
methods \cite{9} , we have obtained the following ansatz:
\be \ba{cccc}
{\cal E}_{Q}(z) = \dys{z^{\frac{(Q+\varphi)^{2}}{4h}} \Pi_{V}(z)} & , &
\Pi_{V}(z) = \dys{\prod_{m=1}^{\infty} (1-z^{m})^{-1}} & , \ea
\label{e7} \ee
where $\varphi = 2h(1-\psi)$ independent of $a$ or $b$ 
(see Eqs.~(\ref{e4a},\ref{e4b})).
Notice that \mbox{${\cal E}_{Q} \neq {\cal E}_{-Q}$}. In the $U_{q}[SU(2)]$
symmetric case the partition functions are very different, one obtains
\cite{8} :
\be
{\cal E}_{Q} = {\cal E}_{-Q} = 
\sum_{j = |Q|}^{\infty} \left[ z^{\frac{(2j+1-h)^{2}}{4h}} -
z^{\frac{(2j+1+h)^{2}}{4h}} \right] \Pi_{V}(z) \; \; .
\label{e8} \ee
In the expressions (\ref{e7}) and (\ref{e8}), $Q$ is an integer or
half-integer. This result is very interesting because from Eq.~(\ref{e7})
we learn that the operator content in the charge sector $Q$ is given
by a single irrep of a shifted $U(1)$ Kac-Moody algebra \cite{7},
the shift $\varphi$ being related to $\psi$. The expression (\ref{e7}) is the
starting point of the Feigin-Fuchs construction \cite{10,11} of irreps
of Virasoro algebras with \mbox{$c\! < \! 1$} starting from irreps with 
\mbox{$c\gl 1$} (this is the central charge of the XXZ chain). This 
construction will allow us to identify the operator content of systems with 
\mbox{$c\! <\! 1$} with free, fixed, and mixed BC. First, we recall
that for \mbox{$c\! <\! 1$} the central charge is quantized:
\be \ba{ccc}
c = 1 - \dys{\frac{6}{m(m+1)}} & (m=3,4,\ldots ) & , \ea
\label{e9} \ee
and for a given $m$, the highest weights $\Delta_{r,s}$ of unitary
irreps are
\be \ba{cccc}
\Delta_{r,s} = \dys{\frac{\left[ (m+1)r - ms \right] ^{2} - 1}{4m(m+1)}} & , &
1 \leq r \leq m-1 \; \; , \; \; 1 \leq s \leq m & . \ea
\label{e9b} \ee
The corresponding character functions are:
\be \ba{lcl}
\chi_{r,s} & = & \Omega_{r,s} - \Omega_{r,-s} \vspace{4mm} \\
\Omega_{r,s} & = & \dys{\sum_{\alpha = - \infty}^{\infty}
z^{\frac{[ 2m(m+1)\alpha + (m+1)r - ms ]^{2} - 1}{4m(m+1)}} \Pi_{V}(z) } 
\; \; . \ea
\label{e10} \ee
In order to apply the Feigin-Fuchs procedure, let us assume that we have 
fixed $h$ and $\varphi$. Instead of considering $E_{0}^{F}(N)$ as ground-state
energy, we take as ground-state energy $E_{0;0}(N)$, i.e. the ground-state
in the charge-zero sector of the Hamiltonian (\ref{e1}) with $\gamma$ ,
$p$ and $p^{\prime}$ fixed. Next, instead of considering charges taking
values in ${\bf Z}$ or ${\bf Z}+1/2$, it is convenient to work 
with charges in ${\bf Z}_{n}$ or ${\bf Z}_{n+1/2}$, respectively 
(the values of $n$ will be fixed later). Thus, instead of Eqs.~(\ref{e6})
and (\ref{e7}) we have
\be \ba{lcl}
\Fquer_{\qquer;i} & = & \dys{\frac{N}{\pi} 
\left( E_{\qquer;i} - E_{0;0}(N) \right) }  \vspace{4mm} \\
{\cal F}_{\qquer}(N,z) & = & 
\dys{\sum_{i} z^{\Fquer_{\qquer;i}}} \vspace{4mm} \\
{\cal F}_{\qquer}(z) & = & \dys{\lim_{N \rightarrow \infty} } 
{\cal F}_{\qquer}(N,z) \; = \;
\dys{\sum_{\alpha = - \infty}^{\infty} 
z^{\frac{(n\alpha + \qquer + \varphi)^{2} - 
\varphi^{2}}{4h}} \Pi_{V}(z) } 
\; \; , \ea 
\label{e11} \ee
where $\qquer\gl 0,1,\ldots,n-1$ for $N$ even and
\mbox{$\qquer\gl \frac{1}{2},\frac{3}{2},\ldots,n-\frac{1}{2}$} 
for $N$ odd.
We now choose
\be \ba{ccc}
h = \dys{\frac{n^{2}}{4m(m+1)}} & , & 
\dys{\varphi = \frac{n}{2m(m+1)}} \; \; , \ea
\label{e12} \ee
that is $\psi\gl 1-1/n$ , and get 
\be
{\cal F}_{\qquer}(z) = \sum_{\alpha = - \infty}^{\infty}
z^{\frac{[2m(m+1)\alpha + 2m(m+1)\qquer/n + 1 ]^{2} - 1}{4m(m+1)}}
\Pi_{V}(z) \; \; .
\label{e13} \ee
Since the finite-size scaling spectrum given by Eq.~(\ref{e7}) stays
positive for any positive $h$, we will assume that we can use the equations
above for \mbox{$0\! <\! h\! <\! \frac{1}{4}$} also (see Eq.~(\ref{e2})).
The four choices of $n$ given in Table~1 divide the domain 
\mbox{$h\! \geq \! \frac{1}{6}$} into four regions which correspond to the
$p\,$-state Potts models, tricritical $p\,$-state Potts models, $O(p)$ models,
and low temperature $O(p)$ models \cite{12}. 
\begin{table}[hbt]
\caption{Definition of the models. The arguments of the cosine functions
are taken positive. The values of $n$ occur in 
Eq.~(\protect\ref{e12}).}
\begin{center}
\begin{tabular}{|l|c|c|}
\hline
& \rom & \\
 & 
$p\,$-state Potts & 
Tricritical $p\,$-state Potts \\
 & 
$0<p\leq 4$ & 
$0<p\leq 4$ \\
& \rom & \\ 
\hline
& \rom  & \\
Definition of the model & 
$\frac{1}{2} \sqrt{p}=\cos \pi (1-\frac{1}{4h})$ &
$\frac{1}{2} \sqrt{p}=\cos \pi (\frac{1}{4h}-1)$ \\
& \rom  & \\
Domain of $h$ &
$\frac{1}{2}\geq h \geq \frac{1}{4}$ &
$\frac{1}{4}\geq h \geq \frac{1}{6}$ \\
& \rom & \\
Values of $n$ &
$m+1$ &
$m$ \\
\rule{0mm}{2mm} & & \\
\hline
\multicolumn{3}{c}{\rule{0mm}{6mm}} \\
\hline
& \rom & \\
 & 
$O(p)$ & Low temperature $O(p)$ \\
 & 
$-2<p\leq 2$ & 
$-2<p\leq 2$ \\ 
& \rom & \\
\hline
& \rom & \\
Definition of the model & 
$\frac{1}{2} p =\cos \pi (\frac{1}{h}-1)$ &
$\frac{1}{2} p =\cos \pi (1-\frac{1}{h})$ \\
& \rom & \\
Domain of $h$ &
$1\geq h \geq \frac{1}{2}$ &
$h \geq 1$ \\
& \rom & \\
Values of $n$ &
$2m$ &
$2(m+1)$ \\
& \rom & \\
\hline
\end{tabular}
\end{center}
\end{table}
Let us first consider the case of the $p\,$-state Potts model 
\mbox{($n\gl m+1$)}.
Comparing Eqs.~(\ref{e10}) and (\ref{e13}), we have
\be
{\cal F}_{-\qquer} = \Omega_{1,2\qquer+1}
\label{e14} \ee
and
\begin{eqnarray}
\chi_{1,2v+1}  = & {\cal F}_{-v} - {\cal F}_{v+1} & 
\hspace*{2cm} \qquer \; \mbox{integer} 
\vspace{4mm} \label{e15} \\
\chi_{1,2v}  = & {\cal F}_{\frac{1}{2}-v} - 
{\cal F}_{\frac{1}{2}+v} & 
\hspace*{2cm} \qquer \; \mbox{half-integer} \; \; . 
\label{e16} \end{eqnarray}
In order to clarify the physical significance of our results, let us take 
\mbox{$m\gl 3$}, \mbox{$n\gl 4$}. We find
\be \ba{lclcl}
\chi_{1,1} & = & {\cal F}_{0} - {\cal F}_{1} & , & \Delta = 0 
\; \; , \vspace{4mm} \\
\chi_{1,3} & = & {\cal F}_{-1} - {\cal F}_{2} & , & \Delta = \dys{\frac{1}{2}} 
\; \; . \ea 
\label{e17} \ee
These are precisely the surface exponents of the Ising model (or, alternatively,
the exponents for fixed BC which are the same) \cite{13,14} .
If we take $\qquer$ half-integer, we have
\be \ba{lclcl}
\chi_{1,2} & = & {\cal F}_{-\frac{1}{2}} - {\cal F}_{\frac{3}{2}} & , &
\Delta = \dys{\frac{1}{16}} \; \; , \ea
\label{e18} \ee
which corresponds to the case of mixed BC \cite{13} .
We now consider \mbox{$m\gl 5$}, \mbox{$n\gl 6$} (the three-state
Potts model) and $\qquer$ integer. We obtain
\be \ba{lclcl}
\chi_{1,1} & = & {\cal F}_{0} - {\cal F}_{1} & , & \Delta = 0 
\; \; , \vspace{4mm} \\
\chi_{1,3} & = & {\cal F}_{-1} - {\cal F}_{2} & , & \Delta = \dys{\frac{2}{3}} 
\; \; , \vspace{4mm} \\
\chi_{1,5} & = & {\cal F}_{-2} - {\cal F}_{3} & , & \Delta = 3 
\; \; . \ea
\label{e19} \ee
These are again the known surface exponents (free BC) or the exponents
corresponding to fixed BC. If we take $\qquer$ half-integer
we find
\be \ba{lclcl}
\chi_{1,2} & = & {\cal F}_{-\frac{1}{2}} - 
{\cal F}_{\frac{3}{2}} & , & \Delta = \dys{\frac{1}{8}} 
\; \; , \vspace{4mm} \\
\chi_{1,4} & = & {\cal F}_{-\frac{3}{2}} - 
{\cal F}_{\frac{5}{2}} & , & \Delta = \dys{\frac{13}{8}} 
\; \; . \ea
\label{e20} \ee
The values \mbox{$\Delta\gl \frac{1}{8}$} and 
\mbox{$\Delta\gl \frac{13}{8}$} are indeed the exponents of the three-state
Potts model with mixed BC \cite{13}. The case \mbox{$m\gl 4$} does not
correspond to the tricritical Ising model. It has the same modular invariant
partition function as the tricritical Ising model but with another distribution
of the primary fields between the charge sectors \cite{1}.
 
From the above considerations, we will assume that for all values of $m$,
Eq.~(\ref{e15}) gives us the surface (or fixed BC) exponents and Eq.~(\ref{e16})
the exponents for mixed BC. The operator contents for free (fixed) and mixed
BC for various models are given in Table~2.
\begin{table}[hbt]
\caption{The operator content for free (fixed) and mixed boundary conditions
and their relation to the XXZ chain for various models. (For the $p\,$-state
Potts case see Eqs.~(\protect\ref{e15}) and (\protect\ref{e16}) ).}
\begin{center}
\begin{tabular}{|l|r@{$\;=\;$}l|r@{$\;=\;$}l|}
\hline
\multicolumn{1}{|c}{\rom} &
\multicolumn{2}{|c|}{\rom} & 
\multicolumn{2}{c|}{\rom} \\
\multicolumn{1}{|c}{\mbox{}} &
\multicolumn{2}{|c|}{$p\,$-state Potts} & 
\multicolumn{2}{c|}{Tricritical $p\,$-state Potts} \\
\multicolumn{1}{|c}{\rom} &
\multicolumn{2}{|c|}{\rom} & 
\multicolumn{2}{c|}{\rom} \\
\hline
\multicolumn{1}{|c}{\rom} &
\multicolumn{2}{|c|}{\rom} & 
\multicolumn{2}{c|}{\rom} \\
 &
${\cal F}_{-\qquer}$ & $\Omega_{1,2\qquer+1}$ &
${\cal F}_{\qquer}$ & $\Omega_{2\qquer+1,1}$ \\
\multicolumn{1}{|c}{\rom} &
\multicolumn{2}{|c|}{\rom} & 
\multicolumn{2}{c|}{\rom} \\
Free (fixed) BC &
$\chi_{1,2v+1}$ & ${\cal F}_{-v} - {\cal F}_{v+1}$ &
$\chi_{2v+1,1}$ & ${\cal F}_{v} - {\cal F}_{-v-1}$ \\
\multicolumn{1}{|c}{\rom} &
\multicolumn{2}{|c|}{\rom} & 
\multicolumn{2}{c|}{\rom}  \\
Mixed BC &
$\chi_{1,2v}$ & ${\cal F}_{-v+\frac{1}{2}} - {\cal F}_{v+\frac{1}{2}}$ &
$\chi_{2v,1}$ & ${\cal F}_{v-\frac{1}{2}} - {\cal F}_{-v-\frac{1}{2}}$ \\
\multicolumn{1}{|c}{\rom} &
\multicolumn{2}{|c|}{\rom} & 
\multicolumn{2}{c|}{\rom}  \\
\hline
\multicolumn{5}{c}{\rule{0mm}{6mm}} \\
\hline
\multicolumn{1}{|c}{\rom} &
\multicolumn{2}{|c|}{\rom} & 
\multicolumn{2}{c|}{\rom} \\
\multicolumn{1}{|c}{\mbox{}} &
\multicolumn{2}{|c|}{$O(p)$} & 
\multicolumn{2}{c|}{Low temperature $O(p)$} \\
\multicolumn{1}{|c}{\rom} &
\multicolumn{2}{|c|}{\rom} & 
\multicolumn{2}{c|}{\rom} \\
\hline
\multicolumn{1}{|c}{\rom} &
\multicolumn{2}{|c|}{\rom} & 
\multicolumn{2}{c|}{\rom}  \\
 &
${\cal F}_{\qquer}$ & $\Omega_{\qquer+1,1}$ &
${\cal F}_{-\qquer}$ & $\Omega_{1,\qquer+1}$ \\
\multicolumn{1}{|c}{\rom} &
\multicolumn{2}{|c|}{\rom} & 
\multicolumn{2}{c|}{\rom}  \\
Free (fixed) BC &
$\chi_{r,1}$ & ${\cal F}_{r-1} - {\cal F}_{-r-1}$ &
$\chi_{1,s}$ & ${\cal F}_{1-s} - {\cal F}_{1+s}$ \\
\multicolumn{1}{|c}{\rom} &
\multicolumn{2}{|c|}{\rom} & 
\multicolumn{2}{c|}{\rom}  \\
\multicolumn{1}{|l}{Mixed BC} & 
\multicolumn{2}{|c|}{m odd only} & 
\multicolumn{2}{c|}{m even only} \\
& 
$\chi_{r,\frac{m-1}{2}}$ & 
${\cal F}_{\frac{m}{2}-1-r} - {\cal F}_{\frac{m}{2}-1+r}$  &
$\chi_{\frac{m}{2}-1,s}$ & 
${\cal F}_{s-\frac{m-1}{2}} - {\cal F}_{-s-\frac{m-1}{2}}$  \\
\multicolumn{1}{|c}{\rom} &
\multicolumn{2}{|c|}{\rom} & 
\multicolumn{2}{c|}{\rom}  \\
\hline
\end{tabular}
\end{center}
\end{table}
Let us derive some known results. For the tricritical Ising model
\mbox{($m\gl n\gl 4$)}, we get
\begin{eqnarray}
\mbox{free BC}  \hspace*{1cm} & 
\left( 0 \right) \oplus \left( \frac{3}{2} \right) &
\vspace{4mm} \label{e21} \\
\mbox{mixed BC}  \hspace*{1cm} & 
\left( \frac{7}{16} \right) & \; \; . 
\label{e22} \end{eqnarray}
Equation~(\ref{e21}) is in agreement with the result of Cardy \cite{13},
Eq.~(\ref{e22}) is new. In the case of the $O(p)$ models, the leading surface
exponent is $\Delta_{2,1}$, again in agreement with Cardy \cite{15} .
 
Let us proceed with the discussion of our results by means of an example.
From the first line of Eq.~(\ref{e19}) we see that combining the finite-size
scaling spectra of the \mbox{$\qquer\gl 0$} and
 \mbox{$\qquer\gl 1$} sectors (see Eq.~(\ref{e11}) ), disregarding all
the doublets (levels which occur in the sectors  \mbox{$\qquer\gl 0$}
and  \mbox{$\qquer\gl 1$}), and keeping only the singlets
(levels which occur in the  \mbox{$\qquer\gl 0$} sector only),
we obtain the vacuum sector of the three-state Potts model. This is the
general structure of the projection mechanism of the finite-size scaling
spectra for \mbox{$c\! <\! 1$} from the spectra for \mbox{$c\gl 1$}.
We can go one step further and check whether the same projection mechanism
also works for a {\em finite number of sites}. For the example of
Eq.~(\ref{e19}) this would imply that all the levels of the sector
\mbox{$\qquer\gl 1$} are exactly degenerate with levels of the sector
\mbox{$\qquer\gl 0$} even for a finite number of sites, and the
remaining levels in the sector \mbox{$\qquer\gl 0$} define the
spectrum of the vacuum sector of the three-state Potts model. The same picture
should be valid for the \mbox{$\qquer\gl -1$} and
\mbox{$\qquer\gl 2$} sectors as well as for the
\mbox{$\qquer\gl -2$} and \mbox{$\qquer\gl 3$}
sectors. 
\begin{table}[p]
\caption{Table of $\Fquer_{\qquer;i}(N)$ values 
(see Eq.~(\protect\ref{e11}) ) for $N\gl 4$, $h\gl \frac{3}{10}$ ,
$\psi\gl \frac{5}{6}$ , and several values of $a$ and $b$ (see 
Eqs.~(\protect\ref{e4a}) and (\protect\ref{e4b}) . The underlined levels
occur in the three-state Potts model with free BC. (The XXZ chain with
$2N$ sites gives the $p\,$-state Potts model with $N$ sites and free BC.)}
\begin{center}
\begin{tabular}{|c|c|c|c|c|c|}
\hline
\multicolumn{2}{|c}{\rom} &
\multicolumn{2}{|c|}{\rom} &
\multicolumn{2}{c|}{\rom} \\
\multicolumn{2}{|c}{$a\gl \frac{2}{3}$} &
\multicolumn{2}{|c}{$a\gl \frac{1}{6}$} &
\multicolumn{2}{|c|}{$a\gl b\gl 0$} \\
\multicolumn{2}{|c}{\rom} &
\multicolumn{2}{|c|}{\rom} &
\multicolumn{2}{c|}{\rom} \\
\hline
$\qquer\gl 0$ &
$\qquer\gl 1$ &
$\qquer\gl 0$ &
$\qquer\gl 1$ &
$\qquer\gl 0$ &
$\qquer\gl 1$ \\
\hline
$\underline{0.000\,000}$ & &
$\underline{0.000\,000}$ & &
$\underline{0.000\,000}$ & \\
$0.869\,482$ &
$0.869\,482$ &
$0.573\,943$ &
$0.573\,943$ &
$0.566\,777$ &
$0.566\,777$ \\
$1.347\,175$ &
$1.347\,175$ &
$1.128\,282$ &
$1.128\,282$ &
$1.124\,431$ &
$1.124\,431$ \\
$\underline{1.407\,619}$ & &
$\underline{1.407\,619}$ & &
$\underline{1.407\,619}$ & \\
$2.022\,885$ &
$2.022\,885$ &
$1.693\,423$ &
$1.693\,423$ &
$1.689\,669$ &
$1.689\,669$ \\
$2.520\,290$ &
$2.520\,290$ &
$1.992\,459$ &
$1.992\,459$ &
$1.980\,841$ &
$1.980\,841$ \\
\hline
$\qquer\gl -1$ &
$\qquer\gl  2$ &
$\qquer\gl -1$ &
$\qquer\gl  2$ &
$\qquer\gl -1$ &
$\qquer\gl  2$ \\
\hline
$\underline{0.471\,151}$ & &
$\underline{0.471\,151}$ & &
$\underline{0.471\,151}$ & \\
$\underline{1.071\,362}$ & &
$\underline{1.071\,362}$ & &
$\underline{1.071\,362}$ & \\
$\underline{1.671\,573}$ & &
$\underline{1.671\,573}$ & &
$\underline{1.671\,573}$ & \\
$2.386\,227$ &
$2.386\,227$ &
$1.928\,985$ &
$1.928\,985$ &
$1.920\,189$ &
$1.920\,189$ \\
\hline
$\qquer\gl -2$ &
$\qquer\gl  3$ &
$\qquer\gl -2$ &
$\qquer\gl  3$ &
$\qquer\gl -2$ &
$\qquer\gl  3$ \\
\hline
$\underline{1.806\,467}$ & &
$\underline{1.806\,467}$ & &
$\underline{1.806\,467}$ & \\
\hline
\multicolumn{6}{c}{\rule{0mm}{6mm}} \\
\hline
\multicolumn{2}{|c}{\rom} &
\multicolumn{2}{|c|}{\rom} &
\multicolumn{2}{c|}{\rom} \\
\multicolumn{2}{|c}{$b\gl 1$} &
\multicolumn{2}{|c}{$b\gl 2$} &
\multicolumn{2}{|c|}{$b\gl \infty$} \\
\multicolumn{2}{|c}{\rom} &
\multicolumn{2}{|c|}{\rom} &
\multicolumn{2}{c|}{\rom} \\
\hline
$\qquer\gl 0$ &
$\qquer\gl 1$ &
$\qquer\gl 0$ &
$\qquer\gl 1$ &
$\qquer\gl 0$ &
$\qquer\gl 1$ \\
\hline
$\underline{0.000\,000}$ & &
$\underline{0.000\,000}$ & &
$\underline{0.000\,000}$ & \\
$0.485\,791$ &
$0.485\,791$ &
$0.471\,833$ &
$0.471\,833$ &
$0.471\,151$ &
$0.471\,151$ \\
$1.079\,782$ &
$1.079\,782$ &
$1.071\,757$ &
$1.071\,757$ &
$1.071\,362$ &
$1.071\,362$ \\
$\underline{1.407\,619}$ & &
$\underline{1.407\,619}$ & &
$\underline{1.407\,619}$ & \\
$1.668\,001$ &
$1.668\,001$ &
$1.671\,294$ &
$1.671\,294$ &
$1.671\,573$ &
$1.671\,573$ \\
$1.838\,081$ &
$1.838\,081$ &
$1.808\,041$ &
$1.808\,041$ &
$1.806\,467$ &
$1.806\,467$ \\
\hline
$\qquer\gl -1$ &
$\qquer\gl  2$ &
$\qquer\gl -1$ &
$\qquer\gl  2$ &
$\qquer\gl -1$ &
$\qquer\gl  2$ \\
\hline
$\underline{0.471\,151}$ & &
$\underline{0.471\,151}$ & &
$\underline{0.471\,151}$ & \\
$\underline{1.071\,362}$ & &
$\underline{1.071\,362}$ & &
$\underline{1.071\,362}$ & \\
$\underline{1.671\,573}$ & &
$\underline{1.671\,573}$ & &
$\underline{1.671\,573}$ & \\
$1.823\,501$ &
$1.823\,501$ &
$1.807\,257$ &
$1.807\,257$ &
$1.806\,467$ &
$1.806\,467$ \\
\hline
$\qquer\gl -2$ &
$\qquer\gl  3$ &
$\qquer\gl -2$ &
$\qquer\gl  3$ &
$\qquer\gl -2$ &
$\qquer\gl  3$ \\
\hline
$\underline{1.806\,467}$ & &
$\underline{1.806\,467}$ & &
$\underline{1.806\,467}$ & \\
\hline
\end{tabular}
\setcounter{page}{6}
\end{center}
\end{table}
In Table~3 we give the values of $\Fquer_{\qquer;i}$
for four sites ($h\gl \frac{3}{10}$ , $\psi\gl \frac{5}{6}$), and
$a\gl \frac{2}{3}$, 
$a\gl \frac{1}{6}$, $a\gl b\gl 0$, $b\gl 1, 2,$ and $\infty$ .
The case \mbox{$b\gl \infty$} corresponds to the $U_{q}[SU(2)]$
symmetric case with \mbox{$q\gl \exp (\frac{i \pi}{6})$} . Leaving aside
for the moment the case \mbox{$b\gl \infty$}, we notice the following
features:
\begin{enumerate}
\item The spectra have doublets and singlets. This suggests a supplementary
symmetry in the Hamiltonian. This symmetry goes away if $h$ and $\varphi$ 
are not ``tuned'' like in Eq.~(\ref{e12}) 
\mbox{($\psi \gl 1 - \frac{\gamma}{\pi}$)}. 
\item The doublets occur (like in Table~3 in the sectors $\,{\cal F}_{0}$
and${\,\cal F}_{1}$ for example) in order to allow the projection mechanism
to occur, but also between other sectors (this can be seen for a larger
number of sites).
\item The doublets ``move'' if we modify the values of the parameters
$a$ and $b$ (like in Table~3) or do not (as one can notice for a larger
number of sites).
\item For larger number of sites (eight for the example presented in
Table~3), the degeneracy of a level can be larger than two,
the levels are however distributed among the various sectors such
that the projection mechanism works.
\item {\em The projection mechanism works even for a finite number of sites!}
The levels left (the singlets) after projecting out the doublets occurring in 
the corresponding charge sectors give precisely the spectrum of the three-state
Potts model with free (fixed) BC (see Table~2 of Ref.~\cite{15}).
\end{enumerate}
 
We now consider the $U_{q}[SU(2)]$ symmetric case \mbox{$b\gl \infty$}.
In this case the multiplets are much larger. Since 
\mbox{$q\gl \exp (\frac{i\pi}{6})$}, the multiplets are $(2j+1)$-dimensional
with \mbox{$j\gl 0,1,2$} for $\qquer$ even and
\mbox{$j\gl \frac{1}{2},\frac{3}{2}$} for $\qquer$ odd. The projection
mechanism is very different in this case (see Ref.~\cite{8}, where a superb
explanation of the projection mechanism is given). One again obtains the
exponents (\ref{e19}) but with different multiplicities:
$\left( 0 \right)$ and $\left( 3 \right)$ occur once, but 
$\left( \frac{2}{3} \right)$ occurs twice.
 
The same pattern occurs for an odd number of sites and for any value of
$m$ as long as \mbox{$n\gl m+1$}. The projection mechanism does {\em not}
work however for the $O(p)$ ($n\gl 2m$) or low temperature $O(p)$
($n\gl 2(m+1)$) models where one does not observe the required doublets
in the spectra. The existence of multiplets for 
\mbox{$\psi \gl 1-\frac{\gamma}{\pi}$} suggesting a discrete non-local
symmetry is one of the big mysteries of our work.
 
The projection mechanism described above can certainly be extended to other
systems since the structure of the character expressions of extended
algebras is always similar to Eq.~(\ref{e10}). The first step in this 
direction in order to derive the surface exponents of the \mbox{$N\gl 1$}
superconformal model was already made \cite{16}.


\begin{thebibliography}{20}\itemsep=0pt

\bibitem{1}
F.~Alcaraz, U.~Grimm, and V.~Rittenberg.
\newblock {\em Nucl. Phys. B316 {\em (1989) 735}}.

\bibitem{2}
A.~Cappelli, C.~Itzykson, and J.-B. Zuber.
\newblock {\em Nucl. Phys. B280 {\em (1987) 445}}.

\bibitem{3}
B.~Nienhuis, E.K. Riedel, and M.~Schick.
\newblock {\em Phys. Rev. B27 {\em (1983) 5625}}.

\bibitem{4}
D.~Baranowski, U.~Grimm, V.~Rittenberg, and G.~\mbox{Sch\"{u}tz}.
\newblock {\em to be published}.

\bibitem{5}
A.~Cappelli.
\newblock {\em Phys. Lett. B185 {\em (1987) 82}}.

\bibitem{6}
F.~Alcaraz, M.~Baake, U.~Grimm, and V.~Rittenberg.
\newblock {\em J. Phys. A: Math. Gen. 22 {\em (1989) L5}}.

\bibitem{7}
M.~Baake, Ph. Christe, and V.~Rittenberg.
\newblock {\em Nucl. Phys. B300 {\em (1988) 637}}.

\bibitem{9}
C.J. Hamer and M.T. Batchelor.
\newblock {\em J. Phys. A21 {\em (1988) L173}}.

\bibitem{8}
V.~Pasquier and H.~Saleur.
\newblock {\em Nucl. Phys. B {\em (1989)} (to be published)}.

\bibitem{10}
B.L. Feigin and K.B. Fuchs.
\newblock {\em Funct. Anal. Appl. 16 {\em (1982) 114}}.

\bibitem{11}
V.S. Dotsenko and V.A. Fateev.
\newblock {\em Nucl. Phys. B240 {\em (1984) 312}}.

\bibitem{12}
B.~Nienhuis.
\newblock {\em J. Stat. Phys. 34 {\em (1984) 731}}.

\bibitem{13}
J.L. Cardy.
\newblock {\em Nucl. Phys. B275 {\em (1986) 200}}.

\bibitem{14}
G.~v.~Gehlen and V.~Rittenberg.
\newblock {\em J. Phys. A: Math. Gen. 19 {\em (1986) L631}}.

\bibitem{15}
G.~v.~Gehlen, V.~Rittenberg, and T.~Vescan.
\newblock {\em J. Phys. A: Math. Gen. 20 {\em (1987) 2577}}.

\bibitem{16}
L.~Mezincescu, R.I. Nepomechie, and V.~Rittenberg.
\newblock {\em {\em (1990) } to be published}.

\end{thebibliography}
\end{document}